\documentclass[aps,prl,twocolumn,groupedaddress,showpacs]{revtex4-1}

\usepackage{amsmath,amssymb,amsthm}
\usepackage{txfonts} 
\usepackage{graphicx}

\begin{document}


\title{Isometric immersions and self-similar buckling in Non-Euclidean elastic sheets.}



\author{John Gemmer}
\email[]{john\_gemmer@brown.edu}
\affiliation{Division of Applied Mathematics, Brown University, Providence, RI 02906, USA}

\author{Eran Sharon}
\email[]{eran@vms.huji.ac.il}
\affiliation{Racah Institute of Physics, The Hebrew University, Jerusalem 91904 Israel}

\author{Shankar Venkataramani}
\email[]{shankar@math.arizona.edu}
\affiliation{Department of Mathematics, University of Arizona, Tucson, AZ 85721, USA}


\date{\today}

\begin{abstract}

The edge of torn elastic sheets and growing leaves often form a hierarchical buckling pattern.  Within non-Euclidean plate theory  this complex morphology  can be understood as low bending energy isometric immersions of hyperbolic Riemannian metrics.  With this motivation we study the isometric immersion problem in  strip and disk geometries. By finding explicit piecewise smooth solutions of hyperbolic Monge-Ampere equations on a strip, we show there exist periodic isometric immersions of hyperbolic surfaces in the small slope regime. We extend these solutions to exact isometric immersions through resummation of a formal asymptotic expansion. In the disc geometry we construct self-similar fractal-like isometric immersions for disks with constant negative curvature. The solutions in both the strip and disc geometry qualitatively resemble the patterns observed experimentally and numerically in torn elastic sheets, leaves and swelling hydrogels. For hyperbolic non-Euclidean sheets, complex wrinkling patterns are thus possible within the class of finite bending energy isometric immersions. Further, our results identify the key role of the degree of differentiability (regularity) of the isometric immersion in determining the global structure of a non-Euclidean elastic sheet in 3-space. 

\end{abstract}


\maketitle


%
%

\begin{figure}[ht]
\begin{center}
\includegraphics[width=.47\textwidth]{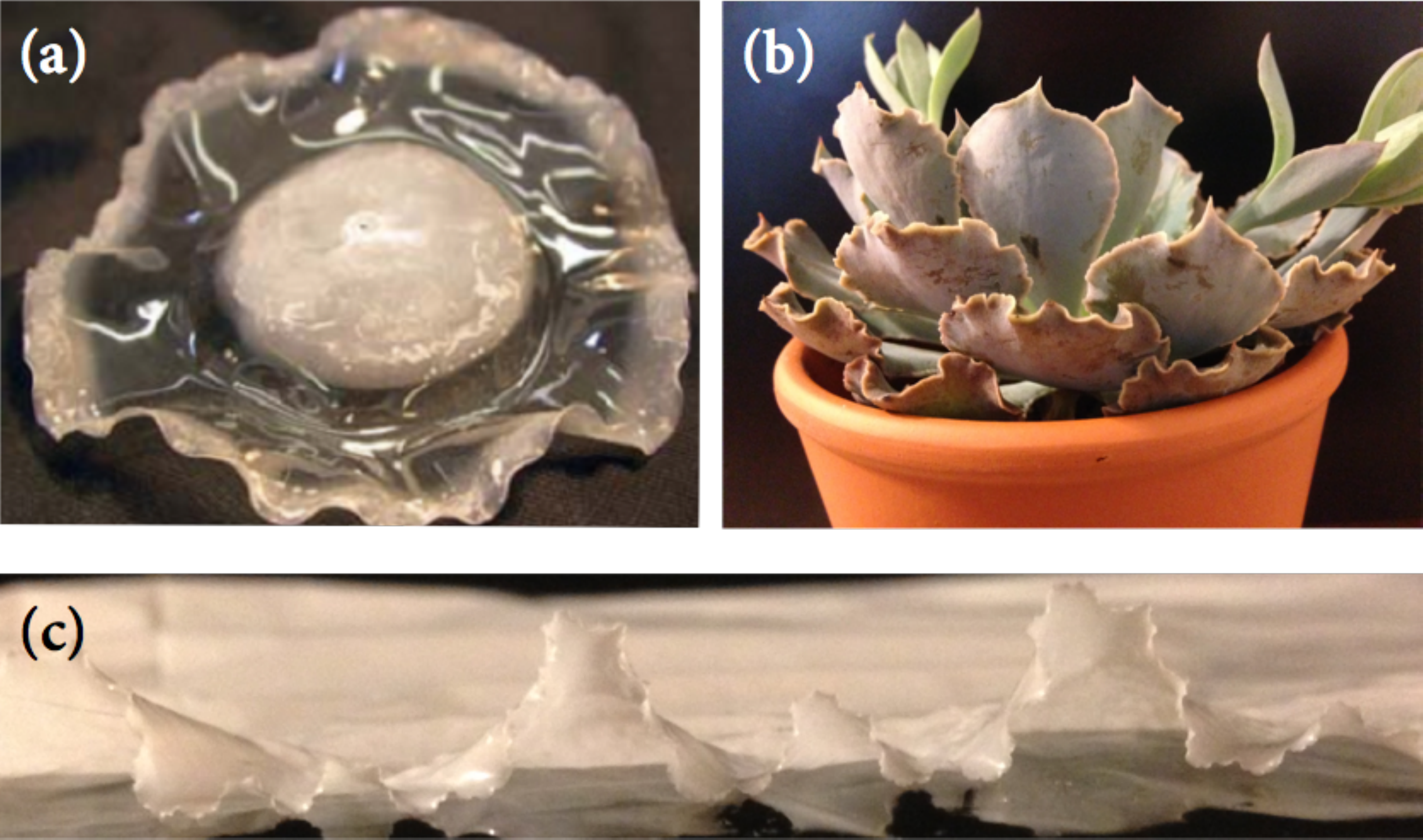}
\caption{Examples of periodic and self-similar wrinkling patterns in swelling thin elastic sheets. (a) Hydrogel disk with non-uniform swelling pattern. (b) Ornamental echeveria house plant. (c) Edge of a torn trash bag. }\label{fig:Examples}
\end{center}
\end{figure}

The rippling patterns observed in torn elastic sheets \cite{sharon2002buckling, audoly2003self, sharon2007geometrically}, leaves \cite{audoly2002ruban, marder2003theory, marder2003shape, marder2006geometry} and swelling hydrogels \cite{efrati2007spontaneous, klein2007shaping, kim2012designing} provide striking examples of periodic and self-similar patterns; see Fig. \ref{fig:Examples}. Within the formalism of finite elasticity, it is understood that such patterns arise from the sheet buckling to relieve growth induced residual strains \cite{goriely2005differential, ben2005growth}. While numerical experiments set within this framework have been able to qualitatively replicate these patterns \cite{marder2003shape, audoly2003self, vetter2013subdivision}, there is no complete theoretical understanding of the mechanism behind the self-similar patterns. On the one hand, such complex, self similar patterns are usually associated with ``strongly frustrated" systems, e.g. elastic sheets with boundary conditions that preclude the possibility of relieving in plane strains \cite{GiOrtz,BCDM, bella2014metric}, or at alloy-alloy interfaces between distinct crystal structures \cite{KM,conti}. On the other hand, many growth patterns generate residual in-plane strains which can be entirely relieved by the sheet forming part of a surface of revolution \cite{marder2006geometry} or a helix \cite{bella2014metric}. Given that generically this system is not strongly frustrated, why then do we observe self-similar buckling patterns? In this Letter we address this puzzle by showing that for a large class of growth profiles there exist periodic and self-similar deformations of the sheet with vanishing in-plane strain. The construction of these surfaces consists of gluing together local solutions of an isometric immersion problem along ``lines of inflection'' and ``branch points'' in such a manner that the resulting surface has finite bending energy. We propose that the sheet introduces these defects to locally reduce large bending content which necessarily results from the extrinsic geometry of the isometric immersions.

One model of swelling thin elastic sheets is the non-Euclidean formalism of elasticity. This model posits that growth permanently deforms the intrinsic distance between material coordinates which is encoded in a Riemannian metric $\mathbf{g}$.  By Gauss's Theorema Egregium, $\mathbf{g}$ generates an intrinsic definition of Gaussian curvature $K$ throughout the sheet \cite{spivakII}  in which points where $K<0$ ($K>0$) correspond to local swelling or growth (shrinkage or atrophy) \cite{dervaux2008morphogenesis}. The realized conformation of the sheet in three dimensional space is then one that minimizes bending energy over all isometric immersions of $\mathbf{g}$ \cite{lewicka2011scaling}. 

\begin{figure}[ht]
\begin{center}
\includegraphics[width=.4\textwidth]{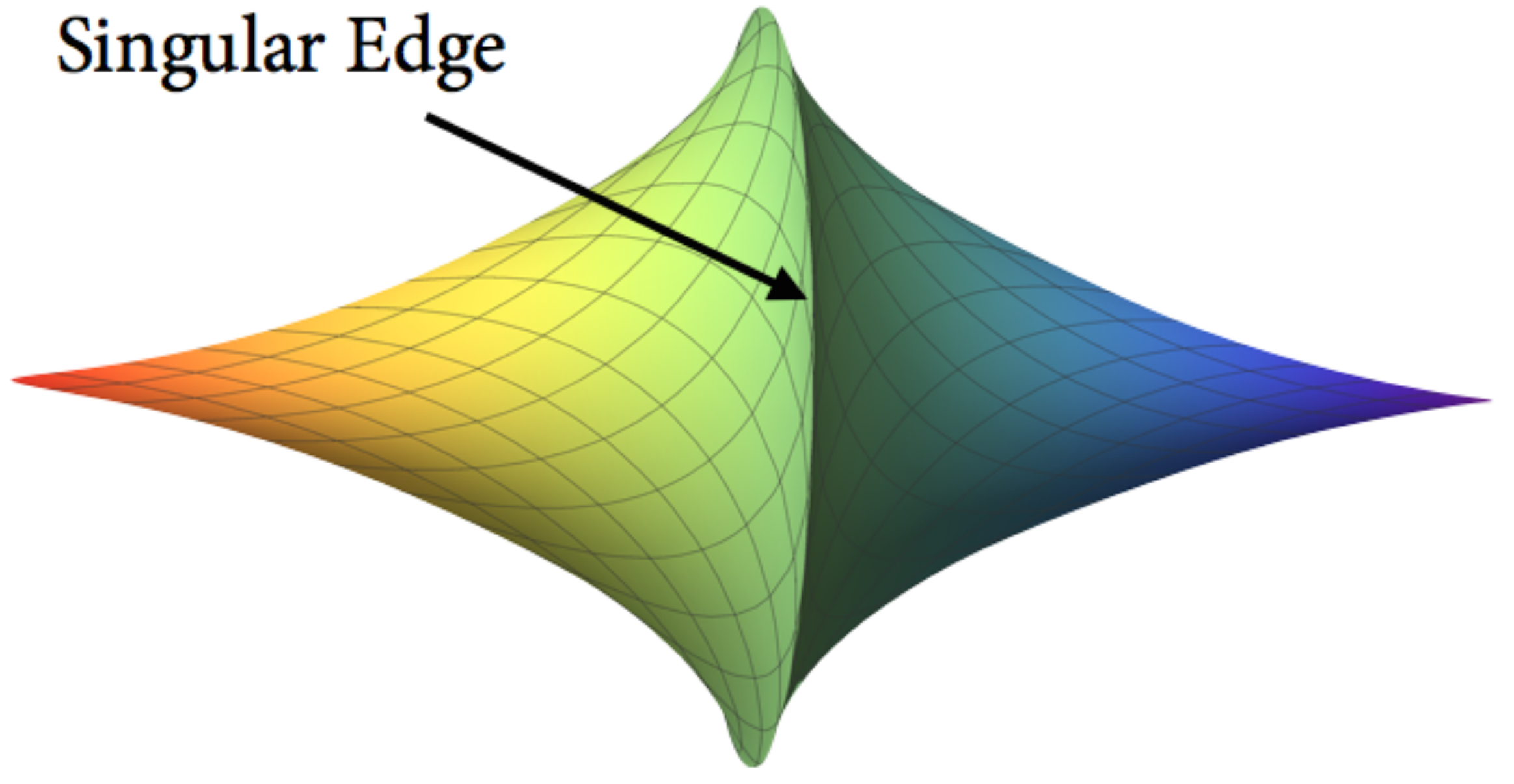}
\caption{Singular rim bounding surfaces of revolution with constant negative Gaussian curvature, e.g the tractricoid (a.k.a ``the'' pseudosphere). Near the singular edge one of the principal curvatures diverges and the bending energy becomes infinite across the curve.} \label{fig:pseudosphere}
\end{center} 
\end{figure}

For $K$ uniformly negative, however,  the extrinsic geometry of the system imposes that with increasing domain size smooth isometries will develop singularities where one of the principal curvatures diverges \cite{efimov1964generation} and for $K=-1$ these singularities form curves -- ``singular edges'' -- \cite{amsler1955surfaces}; see Fig. \ref{fig:pseudosphere}. Across the singular edge the bending energy diverges and hence with increasing domain size a different smooth isometry will be selected by energy minimization. The energetic price paid for switching to another isometry is a global increase of the bending content throughout the sheet; indeed for $K=-1$ the maximum principal curvature of smooth isometries grows exponentially with domain size \cite{gemmer2011shape}. For the case of $K=-1$ it was shown that to mediate the exponentially growing bending content an energetically favorable alternative is for the sheet to adopt the configuration of an $n$-fold periodic ``monkey saddle'' isometry \cite{gemmer2011shape}.  These monkey saddles are not smooth across lines of inflection - every smooth isometry locally has the shape of a saddle - but nevertheless have finite bending energy. However, for a sheet with small but non-zero thickness, it has been shown from force and moment that these defects are in fact smoothed out by boundary layers yielding a smooth local minimizer of the elastic energy \cite{gemmer2012defects}. That is, it is only in the vanishing thickness limit that these local minimizers become non-smooth. The possibly non-smooth isometries selected by the energy in the vanishing thickness limit have recently been coined ``asymptotic isometries'' \cite{chopin2014roadmap}.

To better understand the role the existence of the singular edge plays in the selection of possibly non-smooth isometries we now consider a strip geometry $-\infty< x< \infty$,  $0\leq y<\infty$ with metric
\begin{equation}
\mathbf{g}=(1+\epsilon^2 f(y))\,dx^2+dy^2, \label{eq:StripMetric}
\end{equation}
where $f$ is a function satisfying $f(y)>0$, $f(y)\rightarrow 0$ as $y\rightarrow \infty$, $f^{\prime \prime}(y)>0$ and $\epsilon>0$. This metric corresponds to $y$ dependent growth in the $x$ direction localized near the $y=0$ edge of the sheet and is a generalization of the metrics considered in \cite{marder2003shape, audoly2003self, sharon2007geometrically, liang2009shape, audoly2003self, bella2014metric}. For $\epsilon \ll1$ approximate isometries can be constructed by considering a F\"oppl - von K\'arm\'an ansatz with order $\epsilon^2$ in-plane deformations and order $\epsilon$ out-of-plane deformation $\epsilon w^0(x,y)$. To order $\epsilon^2$ the solvability condition for an isometry is the following small slopes version of Gauss's Theorema Egregium:
\begin{equation}
\det(D^2 (w^0(x,y))=-\frac{f^{\prime \prime}(y)}{2}, \label{eq:mng_amp}
\end{equation}
where $D^2(w^0(x,y))$ denotes the Hessian of $w^0(x,y)$ \cite{audoly2010elasticity}.

\begin{figure}[ht]
 \begin{center}
 \includegraphics[width=.45\textwidth]{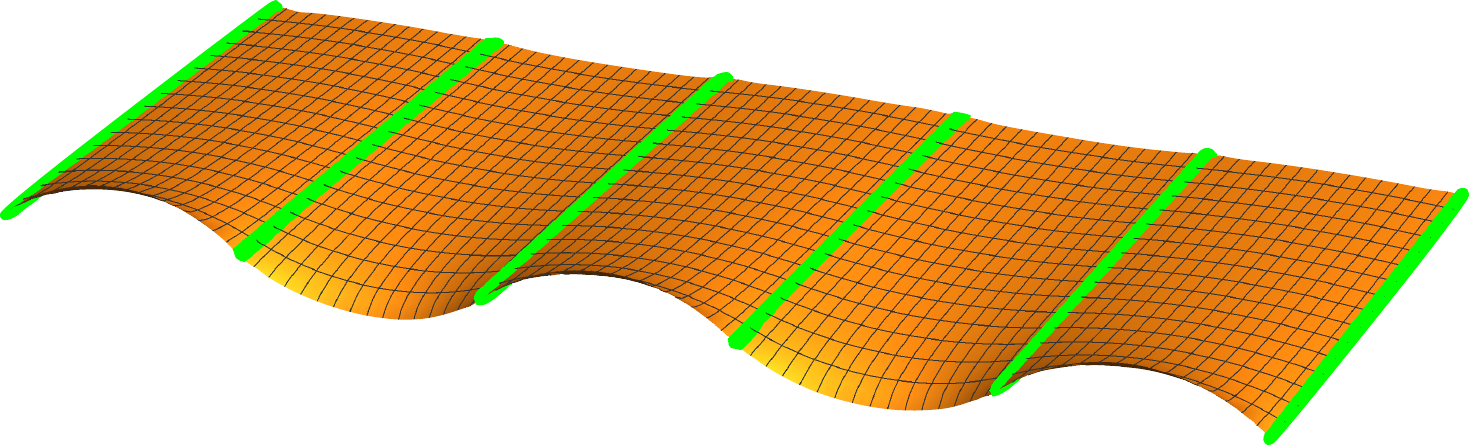}
 \caption{Piecewise smooth small slope isometric immersion with out-of-plane displacement $w^0(x,y)=\psi(x)\phi(y)$ for $\delta=1/2$. The out-of-plane displacement vanishes along the green lines i.e. $\psi(x)=0$, and the periodic surface is created by odd periodically reflecting the smooth surface across these lines. } \label{fig:rippling}
 \end{center}
 \end{figure}

The metric given by Eq. \ref{eq:StripMetric} admits exact smooth isometries in the form of surfaces of revolution \cite{marder2006geometry} or helices \cite{bella2014metric}, however as in the $K=-1$ case they have large bending content throughout the domain. We now seek the analogue of the periodic monkey saddle solutions for this domain. We find that product solutions to  Eq. \eqref{eq:mng_amp} in the form $w^0(x,y)=2^{-\frac{1}{2}}\psi(x)\phi(y)$ exist if for $0<\delta \leq 1$ we take
\begin{align}
4f(y)&=\begin{cases}
2\delta/(1+\delta)\left(1+y/l\right)^{-2\delta/(1-\delta)}& 0 <\delta <1\\
\exp(-2y/l) & \delta =1
\end{cases},
\end{align}
where $l>0$ is the length scale of the swelling and $\psi$ satisfies
\begin{equation}
\psi'^2 \pm k^{2\delta} |\psi|^{2\delta} = 1, \label{eq:mech_energy}
\end{equation}
where $k$ is a constant of integration. We can identify solutions $\psi(x)$ with the motion of a unit mass particle in the potential $V(\psi) =\pm \frac{1}{2} (k|\psi|)^{2\delta}$ and with the positive branch we get periodic solutions with amplitude $1/k$ and half wavelength $\lambda_k\sim k^{-1}$. Note, however, for $\delta \neq 1$ $\psi$ is not smooth across the lines of inflection $\psi=0$; see Fig. \ref{fig:rippling}. Nevertheless, for $1/4<\delta \leq  1$ it follows that $\overline{\psi}^{\prime \prime}$ is square integrable and thus these deformations have finite bending energy. 

\begin{figure}[ht]
\begin{center}
\includegraphics[width=.45\textwidth]{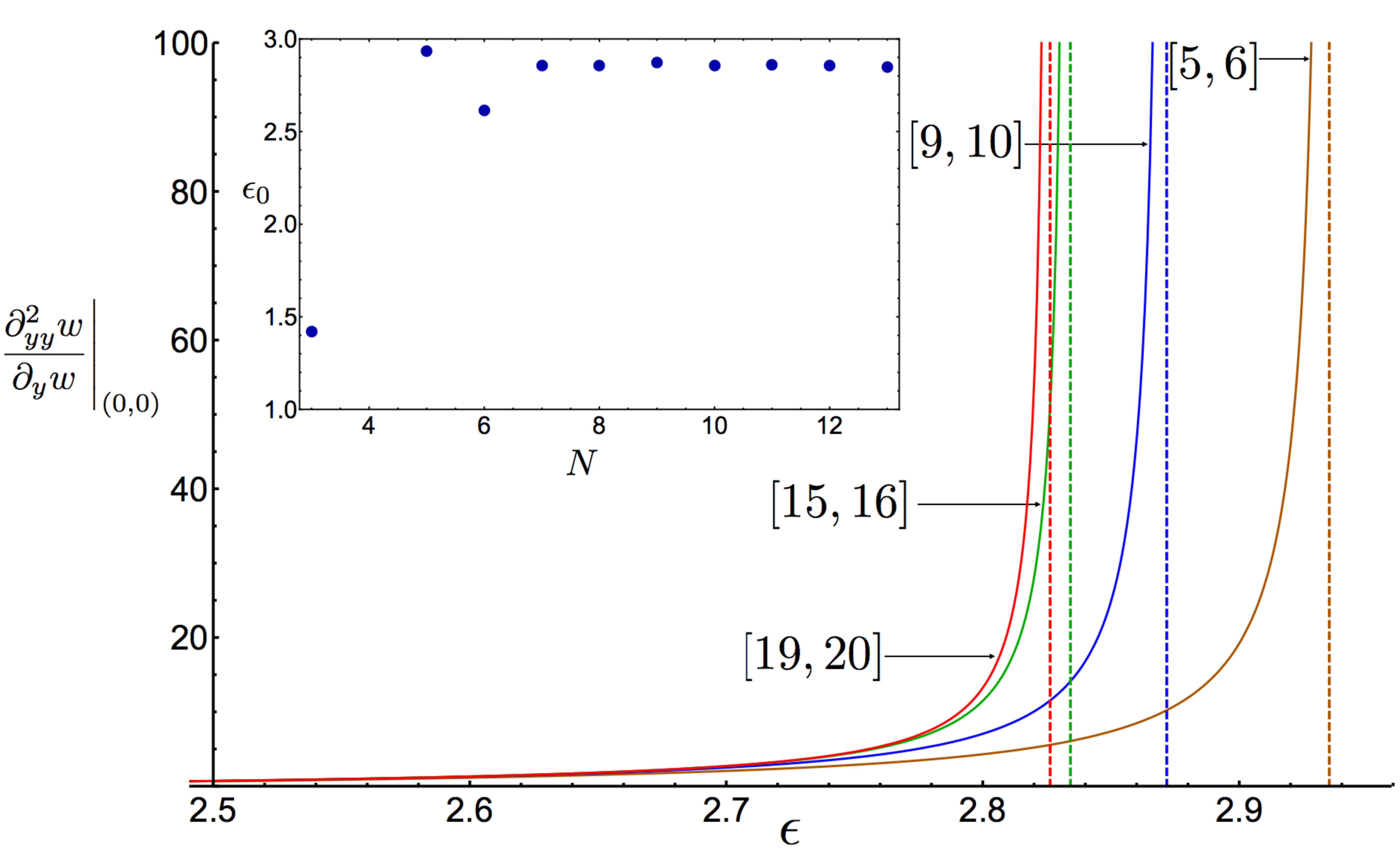}
\end{center}
\caption{$[N,N+1]$ Pad\'{e} approximants to the logarthimic deriviative $\partial^2_{yy}w/\partial_yw$ evaluated at $(0,0)$. The inset figure is a plot of the aproximate value of $\epsilon_0$ determined from the $[N,N+1]$ Pad\'{e} approximants.} \label{fig:DlogPade}
\end{figure}

For $\delta=1$ we now explore the possibility of ``lifting" the small slope isometries to full isometric immersions using the formal asymptotic expansion in $\epsilon$. By solving the equations for an isometry order by order it follows that the order $\epsilon^{2\alpha+1}$ corrections to the out-of-plane displacement $w=\sum_{i=0}^{\infty}\epsilon^{2\alpha+1}w^{\alpha}$ are of the form
\begin{equation*}
w^{\alpha}(x,y)=k^{-1}e^{-(2\alpha+1)y/l}\sum_{n=0}^{\alpha}a_{n,\alpha}\left((kl)^{-1}\right)\cos((2n+1)k x),
\end{equation*}
where $a_{n,i}$ are polynomials. Now, the full series representation $w$ is only asymptotic in $\epsilon$, i.e. it could diverge. Indeed, for increasing values of $\epsilon$ -- equivalently Gaussian curvature -- we expect the surface to develop a  singular edge, with the first singularity appearing at $(0,0)$, i.e. where $\left|\partial^2_{yy}w\right|$ attains its maximum value. If $\partial^2_{yy}w$ has a singularity of the form $\partial^2_{yy}w(0,y)\sim A(y+\epsilon-\epsilon_0)^{-(\beta+1)}$ near $(0,0)$ we can approximate $\epsilon_0$ and $\beta$ by the poles and residues of the Pad\'{e} approximants to the logarithmic derivative $\partial^2_{yy}w(0,0)/\partial_yw(0,0)$ \cite{baker1961application}. In Fig. \ref{fig:DlogPade} we plot the Dlog Pad\'{e} approximants for the case $(kl)=1$ and find that $\epsilon_0\approx 2.86$ and $\beta=.500158$. This strongly indicates that the principal curvature $\kappa_{y}$  scales like $s^{-\frac{1}{2}}$, where $s$ is the arclength measured from the singular edge; a result that is consistent with the singular edge of the pseudosphere and other known hyperbolic surfaces of revolution.

\begin{figure}[ht]
\begin{center}
\includegraphics[width=.45\textwidth]{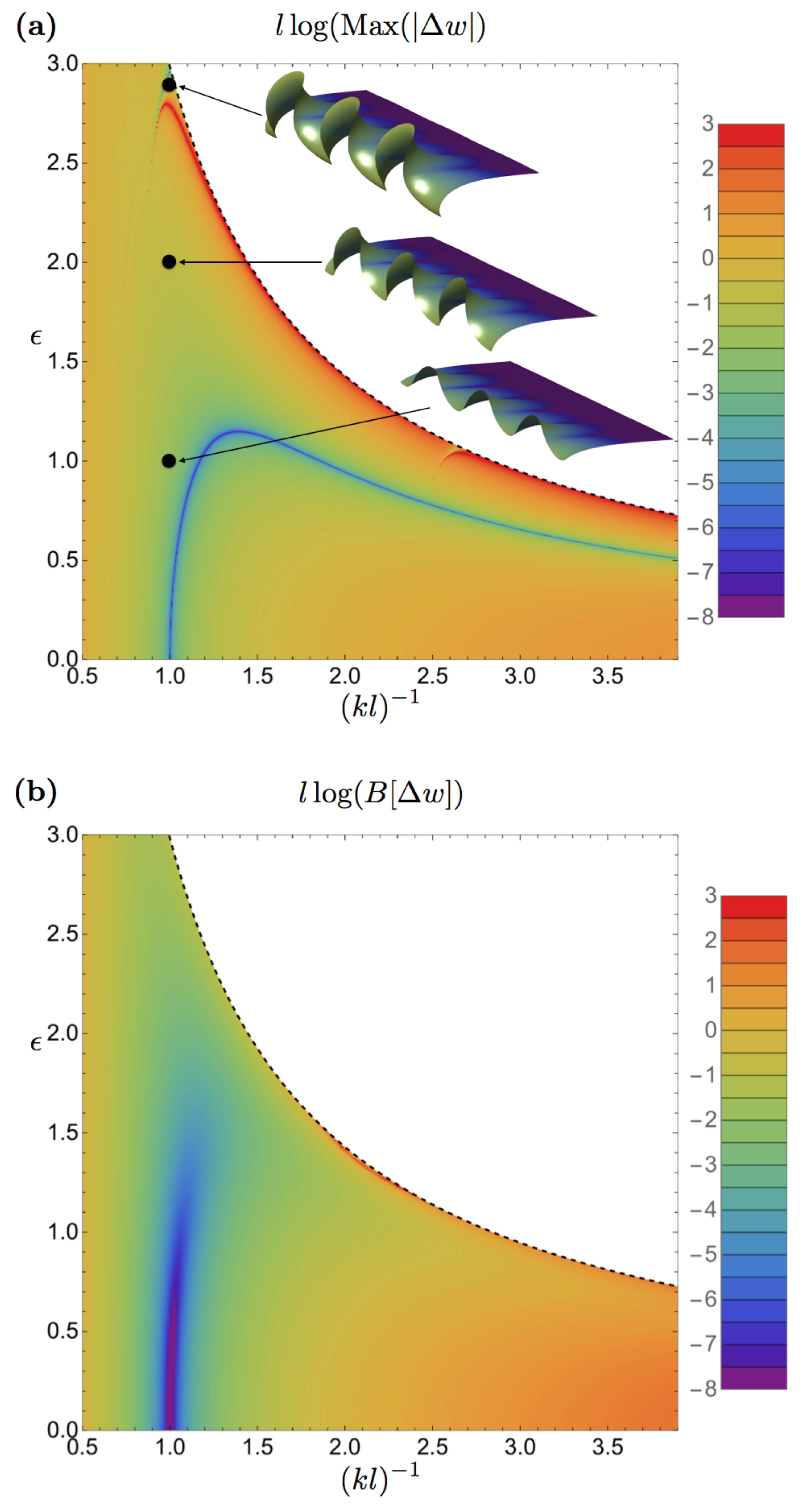}
\end{center}
\caption{(a) Contour plot of the $[9,9]$ Pad\'{e} approximants to $\Delta w$ evaluated at $(0,0)$ as a function of $(kl)^{-1}$ and $\epsilon$ for the specific value $\delta=1$. The dashed curve indicates values at which the singular curve first appears at $(0,0)$. (b) Contour plot of the bending energy $B[\Delta w]$ as a function of $(kl)^{-1}$ and $\epsilon$ for the specific value $\delta=1$ using the $[9,9]$ Pad\'{e} approximants to $\Delta w$. } \label{fig:BendingEnergy}
\end{figure}

The principal curvature $\kappa_{y}$ diverges with increasing $\epsilon$ for all values of $k$ and $l$. This has consequences for modeling non-Euclidean elastic sheets. In particular minimizers of the small slope approximation to the elastic energy do not adequately approximate the full elastic energy.  E.g, with $\delta=1$ and $kl=1$ the solution $w^0(x,y)=k^{-1}e^{-y/l}\cos(kx)$ is harmonic and hence a global minimizer of the bending energy over small slope isometries. However, in Fig. \ref{fig:BendingEnergy}(a) and Fig. \ref{fig:BendingEnergy}(b) we present contour plots of the $[9,9]$ Pad\'{e} approximants to $\Delta w$ evaluated at $(0,0)$ and the following proxy for the bending energy averaged over one half wavelength: $B[\Delta w]=\lambda_k^{-1}\int_{-\lambda_k/2}^{\lambda_k/2}\int_0^{\infty}( \Delta w)^2\,dydx$. Fig. \ref{fig:BendingEnergy}(b) indicates that for moderate values of $\epsilon$, i.e.  $\epsilon \gtrapprox 2$,  the full elastic energy selects shorter wavelengths than predicted by the small slope theory, i.e.  $\epsilon\ll1$. In particular, due to the existence of the singular edge, it follows that for large values of $\epsilon$ the wavelength must satisfy $kl>1$ in order for the isometric immersion to exist. In particular, in a similar manner to the case $K=-1$ to ``go past'' the singular edge it is necessary for this class of periodic isometries to refine their wavelength. The insets in Fig.~\ref{fig:BendingEnergy}(a) also indicate that the series solutions have a unique global wavelength and are smooth up to the singular edge. 

\begin{figure}[ht]
\begin{center}
\includegraphics[width=.4\textwidth]{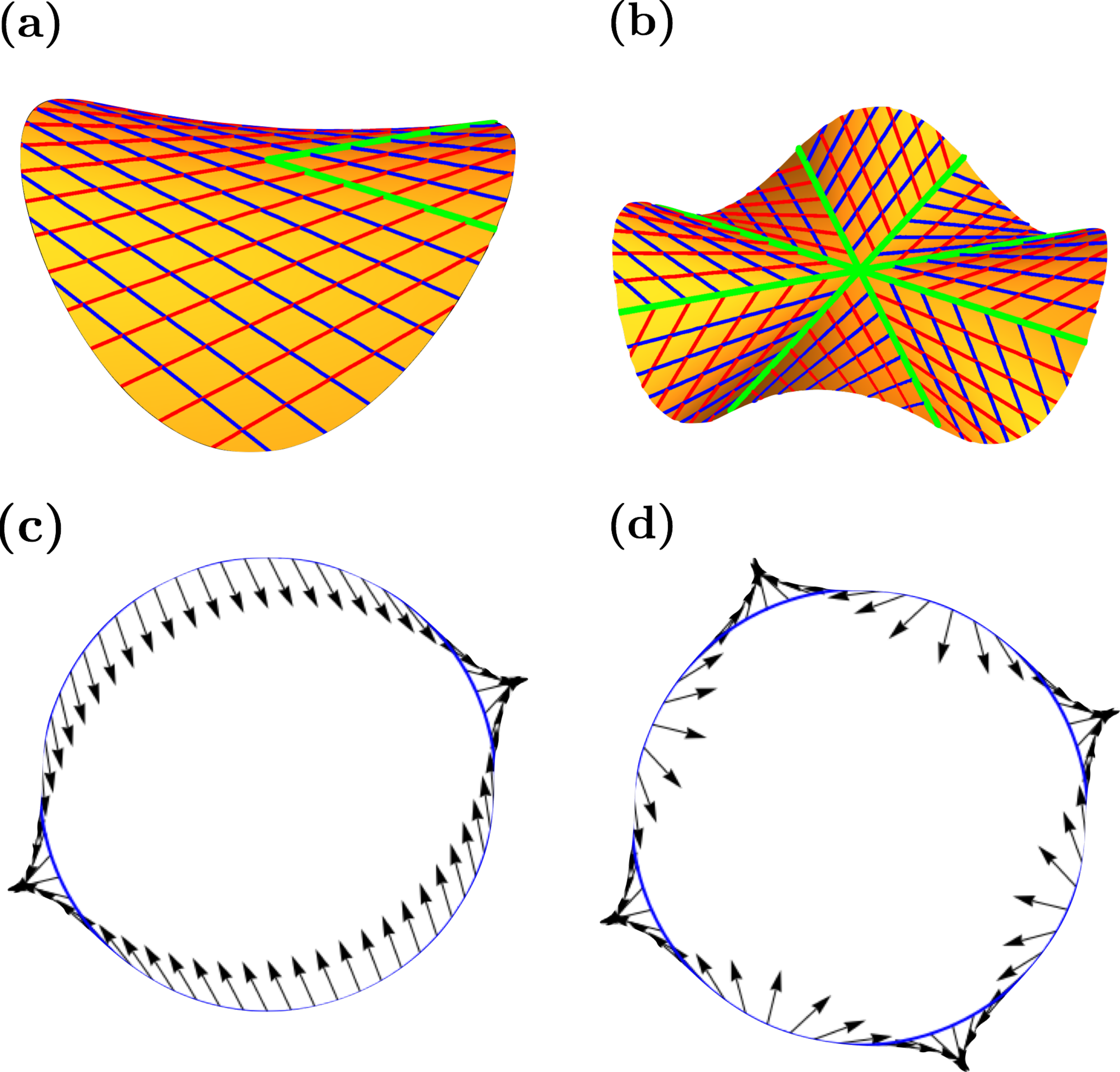}
\end{center}
\caption{(a-b) Small slope isometric immersions $w_4^0(x,y)$ and $\overline{w}_4^0(x,y)$ for constant Gaussian curvature $K=-1$. $\overline{w}_4^0(x,y)$ is constructed by taking odd periodic reflections of the piece of $w_4^0(x,y)$ bounded between the green lines. The mesh on both of these surfaces correspond to their asymptotic lines. (c-d) Direction of the gradient $\nabla w$ along circles centered at the origin. The regular saddle in (a) corresponds to a gradient field with winding number -1, so the gradient map is 1 to 1. The 4-saddle in (b) has winding number -3, so the gradient map is a 3 sheeted covering near the origin. } \label{fig:n-wave}
\end{figure}

We now consider how periodic isometric immersions can introduce further defects in order to lower their bending content. For simplicity we consider solutions to the Monge-Ampere equation $\det(D^2 w^0) = -1$, the small slopes approximation to $K = -1$, on the disk $x^2+y^2\leq R^2$.  In reference \cite{gemmer2011shape},  the n-fold periodic solutions to this equation were formed by taking odd periodic reflections of the solution $w_n^0(x,y)=y(x-\cot(\pi/n)y)$ about the line of inflection $y=\tan(\pi/n)x$; see Fig. \ref{fig:n-wave}(a-b). We adopt the terminology  in \cite{amsler1955surfaces}  and refer to points where lines of inflection intersect -- in this case $x=y=0$ -- as \emph{branch points}. They correspond to branch points of the map $(x,y) \mapsto \nabla w$, or equivalently to branch points for the Gauss normal map \cite{spivak}; see Fig. \ref{fig:n-wave}(c-d).

\begin{figure}[ht]
\includegraphics[width=.45\textwidth]{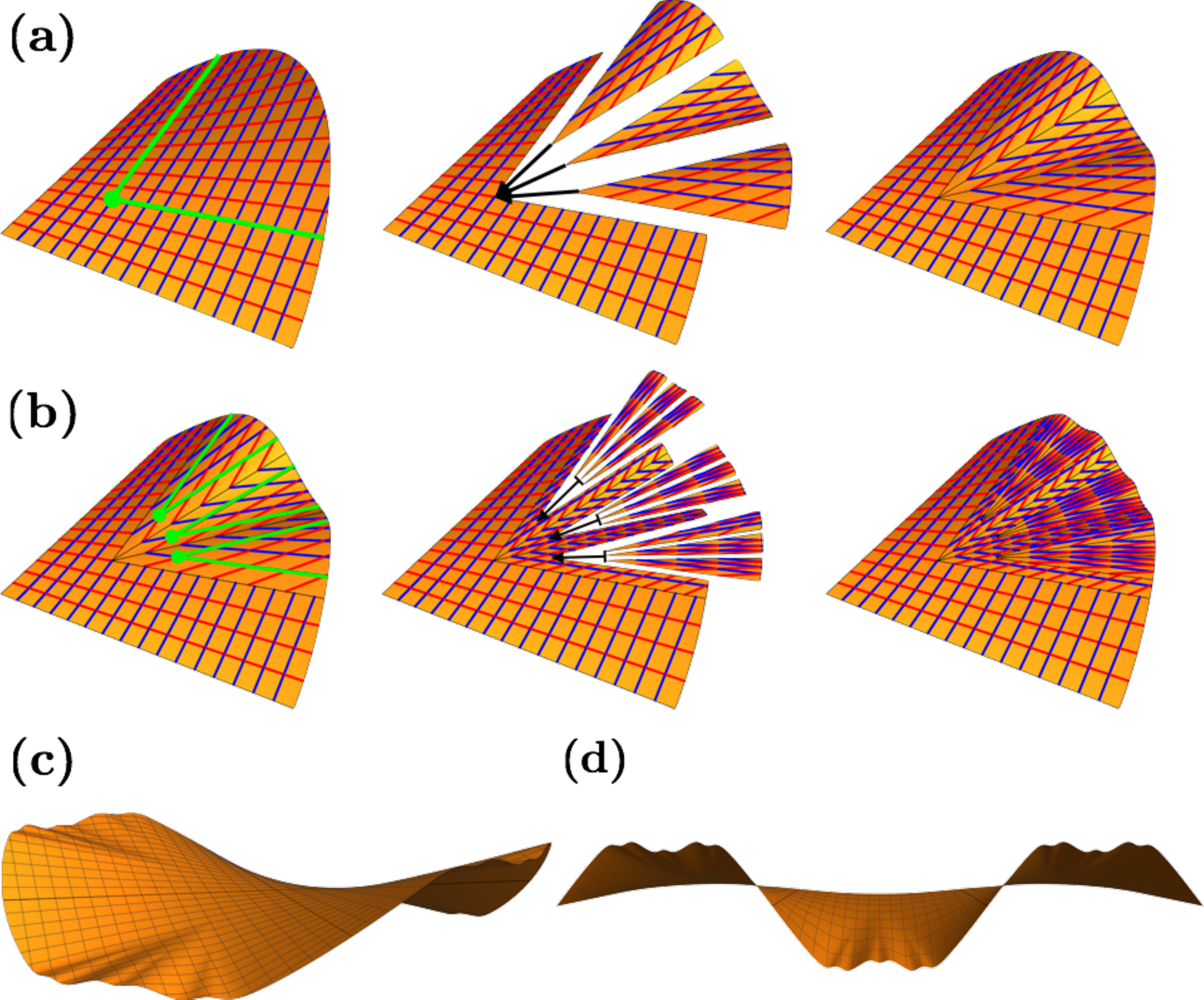}
\caption{Finite bending energy solutions to the Monge-Ampere equation $[w^0,w^0]=-1$. (a) Three subwrinkle solution created by inserting three rotated and translated copies of the solution $w^0_6(x,y)=y(x-\sqrt{3}y)$ onto the solution $w^0_2(x,y)=xy$ at a branch point. (b) Nine subwrinkle solution created by inserting nine copies of $w^0_{12}(x,y)=y(x-(2+\sqrt{3})y)$ at three branch points added onto the three subwrinkle solution. (c) Extension of the nine subwrinkle solution to the full circular domain. (d) The nine subwrinkle solution mapped to the strip geometry by a conformal map.}
\label{fig:SelfSimilar}
\end{figure}

The key to the above construction is that any quadratic surface is ruled by two sets of parallel asymptotic lines and to retain continuity of the tangent plane the lines of inflection are introduced along asymptotic lines; see Fig. \ref{fig:n-wave}(b). Replicating the above process at other points, multiple branch points can be introduced on the surface not just at the origin. For example, for $w_2^0(x,y)=xy$ a  branch point $b_{1,1}$ can be added by removing the sector $x,y\geq 1/\sqrt{2}$ and in this region fitting three appropriately rotated and translated copies of $w_6^0(x,y)=y(x-\sqrt{3}y)$; see Fig \ref{fig:SelfSimilar}(a).  Three more branch points $b_{2,1}$, $b_{2,2}$, $b_{2,3}$ can be added along rays emanating from $b_{1,1}$ that bisect the lines of inflection; see Fig \ref{fig:SelfSimilar}(b). The solution can be extended  by odd periodic reflections to give a small-slopes isometric immersion of the unit disk with $K = -1$.  To illustrate the wrinkling behavior near the edge we map $\overline{w}$ to a strip geometry through a conformal map $h[x+iy] = \overline{w}[e^{x+iy}]$; see Figs. \ref{fig:SelfSimilar}(c-d).

The existence of isometric immersions with branch points also has implications to the modeling of non-Euclidean elastic sheets. As for the strip with $\delta=1$, the solution $w_2^0(x,y)$ is harmonic yet the extension of $w_2^0(x,y)$ to an exact isometric immersion has divergent bending energy for $R\gtrapprox 1.25$ with the bending content concentrated near the singular point $x=y\approx 1.25/\sqrt{2}$ \cite{gemmer2011shape}. We can isometrically immerse disks with larger $R$ by a global refinement of the wavelength i.e taking  $n > 2$. These solutions increase the bending energy globally. An energetically favorable alternative might be to introduce a branch point in the $n = 2$ solution near the singular point, and locally refine the wavelength instead. Indeed, numerics for $\delta=1/3$ in the strip geometry indicate that, even within the small slopes approximation, localized self similar wrinkling profiles may be energetically preferred over global refinement of the wavelength \cite{audoly2003self, vetter2013subdivision}.

The existence of finite bending energy isometric immersions for hyperbolic free sheets ensures that the per unit thickness energy scales like $t^2$. This is in contrast to crumpled sheets which have an energy scale $t^{5/3}$ which is intermediate between the  stretching and bending energies \cite{science.paper,lobkovsky}.  Furthermore, our results show that there are multiple low-energy states as finite bending energy isometric immersions  can be constructed by appropriately gluing together low energy building blocks using lines of inflection and branch points in a variety of ways.  The energy barriers between distinct low energy states are small, $\sim t^2$, and these sheets are thus ``floppy''. Consequently, thin elastic sheets are easily deformed by weak stresses, and the pattern selected may be sensitive to the dynamics of the swelling process, experimental imperfections, or other external forces. A key first step for better understanding the buckling patterns of  hyperbolic free sheets would be to analyze how the optimal elastic energy of the two types of singularities -- lines of inflection and branch points -- scales with the various length scales in the problem: $k^{-1}$, $l$ and $t$.

\begin{acknowledgments}
JG, ES and SV were supported in part by an US-Israel BSF grant 2008432. JG and SV were also supported by the NSF grant DMS-0807501 and JG is currently supported by NSF-RTG grant DMS-1148284.
\end{acknowledgments}

\bibliography{ref}

\def\cprime{$'$}
\begin{thebibliography}{32}%
\makeatletter
\providecommand \@ifxundefined [1]{%
 \@ifx{#1\undefined}
}%
\providecommand \@ifnum [1]{%
 \ifnum #1\expandafter \@firstoftwo
 \else \expandafter \@secondoftwo
 \fi
}%
\providecommand \@ifx [1]{%
 \ifx #1\expandafter \@firstoftwo
 \else \expandafter \@secondoftwo
 \fi
}%
\providecommand \natexlab [1]{#1}%
\providecommand \enquote  [1]{``#1''}%
\providecommand \bibnamefont  [1]{#1}%
\providecommand \bibfnamefont [1]{#1}%
\providecommand \citenamefont [1]{#1}%
\providecommand \href@noop [0]{\@secondoftwo}%
\providecommand \href [0]{\begingroup \@sanitize@url \@href}%
\providecommand \@href[1]{\@@startlink{#1}\@@href}%
\providecommand \@@href[1]{\endgroup#1\@@endlink}%
\providecommand \@sanitize@url [0]{\catcode `\\12\catcode `\$12\catcode
  `\&12\catcode `\#12\catcode `\^12\catcode `\_12\catcode `\%12\relax}%
\providecommand \@@startlink[1]{}%
\providecommand \@@endlink[0]{}%
\providecommand \url  [0]{\begingroup\@sanitize@url \@url }%
\providecommand \@url [1]{\endgroup\@href {#1}{\urlprefix }}%
\providecommand \urlprefix  [0]{URL }%
\providecommand \Eprint [0]{\href }%
\providecommand \doibase [0]{http://dx.doi.org/}%
\providecommand \selectlanguage [0]{\@gobble}%
\providecommand \bibinfo  [0]{\@secondoftwo}%
\providecommand \bibfield  [0]{\@secondoftwo}%
\providecommand \translation [1]{[#1]}%
\providecommand \BibitemOpen [0]{}%
\providecommand \bibitemStop [0]{}%
\providecommand \bibitemNoStop [0]{.\EOS\space}%
\providecommand \EOS [0]{\spacefactor3000\relax}%
\providecommand \BibitemShut  [1]{\csname bibitem#1\endcsname}%
\let\auto@bib@innerbib\@empty
\bibitem [{\citenamefont {Sharon}\ \emph {et~al.}(2002)\citenamefont {Sharon},
  \citenamefont {Roman}, \citenamefont {Marder}, \citenamefont {Shin},\ and\
  \citenamefont {Swinney}}]{sharon2002buckling}%
  \BibitemOpen
  \bibfield  {author} {\bibinfo {author} {\bibfnamefont {E.}~\bibnamefont
  {Sharon}}, \bibinfo {author} {\bibfnamefont {B.}~\bibnamefont {Roman}},
  \bibinfo {author} {\bibfnamefont {M.}~\bibnamefont {Marder}}, \bibinfo
  {author} {\bibfnamefont {G.-S.}\ \bibnamefont {Shin}}, \ and\ \bibinfo
  {author} {\bibfnamefont {H.~L.}\ \bibnamefont {Swinney}},\ }\href@noop {}
  {\bibfield  {journal} {\bibinfo  {journal} {Nature}\ }\textbf {\bibinfo
  {volume} {419}},\ \bibinfo {pages} {579} (\bibinfo {year}
  {2002})}\BibitemShut {NoStop}%
\bibitem [{\citenamefont {Audoly}\ and\ \citenamefont
  {Boudaoud}(2003)}]{audoly2003self}%
  \BibitemOpen
  \bibfield  {author} {\bibinfo {author} {\bibfnamefont {B.}~\bibnamefont
  {Audoly}}\ and\ \bibinfo {author} {\bibfnamefont {A.}~\bibnamefont
  {Boudaoud}},\ }\href@noop {} {\bibfield  {journal} {\bibinfo  {journal}
  {Physical review letters}\ }\textbf {\bibinfo {volume} {91}},\ \bibinfo
  {pages} {086105} (\bibinfo {year} {2003})}\BibitemShut {NoStop}%
\bibitem [{\citenamefont {Sharon}\ \emph {et~al.}(2007)\citenamefont {Sharon},
  \citenamefont {Roman},\ and\ \citenamefont
  {Swinney}}]{sharon2007geometrically}%
  \BibitemOpen
  \bibfield  {author} {\bibinfo {author} {\bibfnamefont {E.}~\bibnamefont
  {Sharon}}, \bibinfo {author} {\bibfnamefont {B.}~\bibnamefont {Roman}}, \
  and\ \bibinfo {author} {\bibfnamefont {H.~L.}\ \bibnamefont {Swinney}},\
  }\href@noop {} {\bibfield  {journal} {\bibinfo  {journal} {Physical Review
  E}\ }\textbf {\bibinfo {volume} {75}},\ \bibinfo {pages} {046211} (\bibinfo
  {year} {2007})}\BibitemShut {NoStop}%
\bibitem [{\citenamefont {Audoly}\ and\ \citenamefont
  {Boudaoud}(2002)}]{audoly2002ruban}%
  \BibitemOpen
  \bibfield  {author} {\bibinfo {author} {\bibfnamefont {B.}~\bibnamefont
  {Audoly}}\ and\ \bibinfo {author} {\bibfnamefont {A.}~\bibnamefont
  {Boudaoud}},\ }\href@noop {} {\bibfield  {journal} {\bibinfo  {journal}
  {Comptes Rendus Mecanique}\ }\textbf {\bibinfo {volume} {330}},\ \bibinfo
  {pages} {831} (\bibinfo {year} {2002})}\BibitemShut {NoStop}%
\bibitem [{\citenamefont {Marder}\ \emph {et~al.}(2003)\citenamefont {Marder},
  \citenamefont {Sharon}, \citenamefont {Smith},\ and\ \citenamefont
  {Roman}}]{marder2003theory}%
  \BibitemOpen
  \bibfield  {author} {\bibinfo {author} {\bibfnamefont {M.}~\bibnamefont
  {Marder}}, \bibinfo {author} {\bibfnamefont {E.}~\bibnamefont {Sharon}},
  \bibinfo {author} {\bibfnamefont {S.}~\bibnamefont {Smith}}, \ and\ \bibinfo
  {author} {\bibfnamefont {B.}~\bibnamefont {Roman}},\ }\href@noop {}
  {\bibfield  {journal} {\bibinfo  {journal} {EPL (Europhysics Letters)}\
  }\textbf {\bibinfo {volume} {62}},\ \bibinfo {pages} {498} (\bibinfo {year}
  {2003})}\BibitemShut {NoStop}%
\bibitem [{\citenamefont {Marder}(2003)}]{marder2003shape}%
  \BibitemOpen
  \bibfield  {author} {\bibinfo {author} {\bibfnamefont {M.}~\bibnamefont
  {Marder}},\ }\href@noop {} {\bibfield  {journal} {\bibinfo  {journal}
  {Foundations of Physics}\ }\textbf {\bibinfo {volume} {33}},\ \bibinfo
  {pages} {1743} (\bibinfo {year} {2003})}\BibitemShut {NoStop}%
\bibitem [{\citenamefont {Marder}\ and\ \citenamefont
  {Papanicolaou}(2006)}]{marder2006geometry}%
  \BibitemOpen
  \bibfield  {author} {\bibinfo {author} {\bibfnamefont {M.}~\bibnamefont
  {Marder}}\ and\ \bibinfo {author} {\bibfnamefont {N.}~\bibnamefont
  {Papanicolaou}},\ }\href@noop {} {\bibfield  {journal} {\bibinfo  {journal}
  {Journal of statistical physics}\ }\textbf {\bibinfo {volume} {125}},\
  \bibinfo {pages} {1065} (\bibinfo {year} {2006})}\BibitemShut {NoStop}%
\bibitem [{\citenamefont {Efrati}\ \emph {et~al.}(2007)\citenamefont {Efrati},
  \citenamefont {Klein}, \citenamefont {Aharoni},\ and\ \citenamefont
  {Sharon}}]{efrati2007spontaneous}%
  \BibitemOpen
  \bibfield  {author} {\bibinfo {author} {\bibfnamefont {E.}~\bibnamefont
  {Efrati}}, \bibinfo {author} {\bibfnamefont {Y.}~\bibnamefont {Klein}},
  \bibinfo {author} {\bibfnamefont {H.}~\bibnamefont {Aharoni}}, \ and\
  \bibinfo {author} {\bibfnamefont {E.}~\bibnamefont {Sharon}},\ }\href@noop {}
  {\bibfield  {journal} {\bibinfo  {journal} {Physica D: Nonlinear Phenomena}\
  }\textbf {\bibinfo {volume} {235}},\ \bibinfo {pages} {29} (\bibinfo {year}
  {2007})}\BibitemShut {NoStop}%
\bibitem [{\citenamefont {Klein}\ \emph {et~al.}(2007)\citenamefont {Klein},
  \citenamefont {Efrati},\ and\ \citenamefont {Sharon}}]{klein2007shaping}%
  \BibitemOpen
  \bibfield  {author} {\bibinfo {author} {\bibfnamefont {Y.}~\bibnamefont
  {Klein}}, \bibinfo {author} {\bibfnamefont {E.}~\bibnamefont {Efrati}}, \
  and\ \bibinfo {author} {\bibfnamefont {E.}~\bibnamefont {Sharon}},\
  }\href@noop {} {\bibfield  {journal} {\bibinfo  {journal} {Science}\ }\textbf
  {\bibinfo {volume} {315}},\ \bibinfo {pages} {1116} (\bibinfo {year}
  {2007})}\BibitemShut {NoStop}%
\bibitem [{\citenamefont {Kim}\ \emph {et~al.}(2012)\citenamefont {Kim},
  \citenamefont {Hanna}, \citenamefont {Byun}, \citenamefont {Santangelo},\
  and\ \citenamefont {Hayward}}]{kim2012designing}%
  \BibitemOpen
  \bibfield  {author} {\bibinfo {author} {\bibfnamefont {J.}~\bibnamefont
  {Kim}}, \bibinfo {author} {\bibfnamefont {J.~A.}\ \bibnamefont {Hanna}},
  \bibinfo {author} {\bibfnamefont {M.}~\bibnamefont {Byun}}, \bibinfo {author}
  {\bibfnamefont {C.~D.}\ \bibnamefont {Santangelo}}, \ and\ \bibinfo {author}
  {\bibfnamefont {R.~C.}\ \bibnamefont {Hayward}},\ }\href@noop {} {\bibfield
  {journal} {\bibinfo  {journal} {Science}\ }\textbf {\bibinfo {volume}
  {335}},\ \bibinfo {pages} {1201} (\bibinfo {year} {2012})}\BibitemShut
  {NoStop}%
\bibitem [{\citenamefont {Goriely}\ and\ \citenamefont
  {Amar}(2005)}]{goriely2005differential}%
  \BibitemOpen
  \bibfield  {author} {\bibinfo {author} {\bibfnamefont {A.}~\bibnamefont
  {Goriely}}\ and\ \bibinfo {author} {\bibfnamefont {M.~B.}\ \bibnamefont
  {Amar}},\ }\href@noop {} {\bibfield  {journal} {\bibinfo  {journal} {Physical
  review letters}\ }\textbf {\bibinfo {volume} {94}},\ \bibinfo {pages}
  {198103} (\bibinfo {year} {2005})}\BibitemShut {NoStop}%
\bibitem [{\citenamefont {Ben~Amar}\ and\ \citenamefont
  {Goriely}(2005)}]{ben2005growth}%
  \BibitemOpen
  \bibfield  {author} {\bibinfo {author} {\bibfnamefont {M.}~\bibnamefont
  {Ben~Amar}}\ and\ \bibinfo {author} {\bibfnamefont {A.}~\bibnamefont
  {Goriely}},\ }\href@noop {} {\bibfield  {journal} {\bibinfo  {journal}
  {Journal of the Mechanics and Physics of Solids}\ }\textbf {\bibinfo {volume}
  {53}},\ \bibinfo {pages} {2284} (\bibinfo {year} {2005})}\BibitemShut
  {NoStop}%
\bibitem [{\citenamefont {Vetter}\ \emph {et~al.}(2013)\citenamefont {Vetter},
  \citenamefont {Stoop}, \citenamefont {Jenni}, \citenamefont {Wittel},\ and\
  \citenamefont {Herrmann}}]{vetter2013subdivision}%
  \BibitemOpen
  \bibfield  {author} {\bibinfo {author} {\bibfnamefont {R.}~\bibnamefont
  {Vetter}}, \bibinfo {author} {\bibfnamefont {N.}~\bibnamefont {Stoop}},
  \bibinfo {author} {\bibfnamefont {T.}~\bibnamefont {Jenni}}, \bibinfo
  {author} {\bibfnamefont {F.~K.}\ \bibnamefont {Wittel}}, \ and\ \bibinfo
  {author} {\bibfnamefont {H.~J.}\ \bibnamefont {Herrmann}},\ }\href@noop {}
  {\bibfield  {journal} {\bibinfo  {journal} {International Journal for
  Numerical Methods in Engineering}\ }\textbf {\bibinfo {volume} {95}},\
  \bibinfo {pages} {791} (\bibinfo {year} {2013})}\BibitemShut {NoStop}%
\bibitem [{\citenamefont {Ortiz}\ and\ \citenamefont {Gioia}(1994)}]{GiOrtz}%
  \BibitemOpen
  \bibfield  {author} {\bibinfo {author} {\bibfnamefont {M.}~\bibnamefont
  {Ortiz}}\ and\ \bibinfo {author} {\bibfnamefont {G.}~\bibnamefont {Gioia}},\
  }\href@noop {} {\bibfield  {journal} {\bibinfo  {journal} {J. Mech. Phys.
  Solids}\ }\textbf {\bibinfo {volume} {42}},\ \bibinfo {pages} {531} (\bibinfo
  {year} {1994})}\BibitemShut {NoStop}%
\bibitem [{\citenamefont {Ben~Belgacem}\ \emph {et~al.}(2000)\citenamefont
  {Ben~Belgacem}, \citenamefont {Conti}, \citenamefont {DeSimone},\ and\
  \citenamefont {M{\"u}ller}}]{BCDM}%
  \BibitemOpen
  \bibfield  {author} {\bibinfo {author} {\bibfnamefont {H.}~\bibnamefont
  {Ben~Belgacem}}, \bibinfo {author} {\bibfnamefont {S.}~\bibnamefont {Conti}},
  \bibinfo {author} {\bibfnamefont {A.}~\bibnamefont {DeSimone}}, \ and\
  \bibinfo {author} {\bibfnamefont {S.}~\bibnamefont {M{\"u}ller}},\
  }\href@noop {} {\bibfield  {journal} {\bibinfo  {journal} {J. Nonlinear
  Sci.}\ }\textbf {\bibinfo {volume} {10}},\ \bibinfo {pages} {661} (\bibinfo
  {year} {2000})}\BibitemShut {NoStop}%
\bibitem [{\citenamefont {Bella}\ and\ \citenamefont
  {Kohn}(2014)}]{bella2014metric}%
  \BibitemOpen
  \bibfield  {author} {\bibinfo {author} {\bibfnamefont {P.}~\bibnamefont
  {Bella}}\ and\ \bibinfo {author} {\bibfnamefont {R.~V.}\ \bibnamefont
  {Kohn}},\ }\href@noop {} {\bibfield  {journal} {\bibinfo  {journal} {Journal
  of Nonlinear Science}\ }\textbf {\bibinfo {volume} {24}},\ \bibinfo {pages}
  {1147} (\bibinfo {year} {2014})}\BibitemShut {NoStop}%
\bibitem [{\citenamefont {Kohn}\ and\ \citenamefont {M{\"u}ller}(1994)}]{KM}%
  \BibitemOpen
  \bibfield  {author} {\bibinfo {author} {\bibfnamefont {R.~V.}\ \bibnamefont
  {Kohn}}\ and\ \bibinfo {author} {\bibfnamefont {S.}~\bibnamefont
  {M{\"u}ller}},\ }\href@noop {} {\bibfield  {journal} {\bibinfo  {journal}
  {Comm. Pure Appl. Math.}\ }\textbf {\bibinfo {volume} {47}},\ \bibinfo
  {pages} {405} (\bibinfo {year} {1994})}\BibitemShut {NoStop}%
\bibitem [{\citenamefont {Conti}(2000)}]{conti}%
  \BibitemOpen
  \bibfield  {author} {\bibinfo {author} {\bibfnamefont {S.}~\bibnamefont
  {Conti}},\ }\href@noop {} {\bibfield  {journal} {\bibinfo  {journal} {Comm.
  Pure Appl. Math.}\ }\textbf {\bibinfo {volume} {53}},\ \bibinfo {pages}
  {1448} (\bibinfo {year} {2000})}\BibitemShut {NoStop}%
\bibitem [{\citenamefont {Spivak}(1979{\natexlab{a}})}]{spivakII}%
  \BibitemOpen
  \bibfield  {author} {\bibinfo {author} {\bibfnamefont {M.}~\bibnamefont
  {Spivak}},\ }\href@noop {} {\emph {\bibinfo {title} {A comprehensive
  introduction to differential geometry. {V}ol. {II}}}},\ \bibinfo {edition}
  {2nd}\ ed.\ (\bibinfo  {publisher} {Publish or Perish, Inc., Wilmington,
  Del.},\ \bibinfo {year} {1979})\ pp.\ \bibinfo {pages} {xv+423}\BibitemShut
  {NoStop}%
\bibitem [{\citenamefont {Dervaux}\ and\ \citenamefont
  {Amar}(2008)}]{dervaux2008morphogenesis}%
  \BibitemOpen
  \bibfield  {author} {\bibinfo {author} {\bibfnamefont {J.}~\bibnamefont
  {Dervaux}}\ and\ \bibinfo {author} {\bibfnamefont {M.~B.}\ \bibnamefont
  {Amar}},\ }\href@noop {} {\bibfield  {journal} {\bibinfo  {journal} {Physical
  review letters}\ }\textbf {\bibinfo {volume} {101}},\ \bibinfo {pages}
  {068101} (\bibinfo {year} {2008})}\BibitemShut {NoStop}%
\bibitem [{\citenamefont {Lewicka}\ and\ \citenamefont
  {Reza~Pakzad}(2011)}]{lewicka2011scaling}%
  \BibitemOpen
  \bibfield  {author} {\bibinfo {author} {\bibfnamefont {M.}~\bibnamefont
  {Lewicka}}\ and\ \bibinfo {author} {\bibfnamefont {M.}~\bibnamefont
  {Reza~Pakzad}},\ }\href@noop {} {\bibfield  {journal} {\bibinfo  {journal}
  {ESAIM: Control, Optimisation and Calculus of Variations}\ }\textbf {\bibinfo
  {volume} {17}},\ \bibinfo {pages} {1158} (\bibinfo {year}
  {2011})}\BibitemShut {NoStop}%
\bibitem [{\citenamefont {Efimov}(1964)}]{efimov1964generation}%
  \BibitemOpen
  \bibfield  {author} {\bibinfo {author} {\bibfnamefont {N.~V.}\ \bibnamefont
  {Efimov}},\ }\href@noop {} {\bibfield  {journal} {\bibinfo  {journal}
  {Matematicheskii Sbornik}\ }\textbf {\bibinfo {volume} {106}},\ \bibinfo
  {pages} {286} (\bibinfo {year} {1964})}\BibitemShut {NoStop}%
\bibitem [{\citenamefont {Amsler}(1955)}]{amsler1955surfaces}%
  \BibitemOpen
  \bibfield  {author} {\bibinfo {author} {\bibfnamefont {M.-H.}\ \bibnamefont
  {Amsler}},\ }\href@noop {} {\bibfield  {journal} {\bibinfo  {journal}
  {Mathematische Annalen}\ }\textbf {\bibinfo {volume} {130}},\ \bibinfo
  {pages} {234} (\bibinfo {year} {1955})}\BibitemShut {NoStop}%
\bibitem [{\citenamefont {Gemmer}\ and\ \citenamefont
  {Venkataramani}(2011)}]{gemmer2011shape}%
  \BibitemOpen
  \bibfield  {author} {\bibinfo {author} {\bibfnamefont {J.~A.}\ \bibnamefont
  {Gemmer}}\ and\ \bibinfo {author} {\bibfnamefont {S.~C.}\ \bibnamefont
  {Venkataramani}},\ }\href@noop {} {\bibfield  {journal} {\bibinfo  {journal}
  {Physica D: Nonlinear Phenomena}\ }\textbf {\bibinfo {volume} {240}},\
  \bibinfo {pages} {1536} (\bibinfo {year} {2011})}\BibitemShut {NoStop}%
\bibitem [{\citenamefont {Gemmer}\ and\ \citenamefont
  {Venkataramani}(2012)}]{gemmer2012defects}%
  \BibitemOpen
  \bibfield  {author} {\bibinfo {author} {\bibfnamefont {J.}~\bibnamefont
  {Gemmer}}\ and\ \bibinfo {author} {\bibfnamefont {S.}~\bibnamefont
  {Venkataramani}},\ }\href@noop {} {\bibfield  {journal} {\bibinfo  {journal}
  {Nonlinearity}\ }\textbf {\bibinfo {volume} {25}},\ \bibinfo {pages} {3553}
  (\bibinfo {year} {2012})}\BibitemShut {NoStop}%
\bibitem [{\citenamefont {Chopin}\ \emph {et~al.}(2014)\citenamefont {Chopin},
  \citenamefont {D{\'e}mery},\ and\ \citenamefont
  {Davidovitch}}]{chopin2014roadmap}%
  \BibitemOpen
  \bibfield  {author} {\bibinfo {author} {\bibfnamefont {J.}~\bibnamefont
  {Chopin}}, \bibinfo {author} {\bibfnamefont {V.}~\bibnamefont {D{\'e}mery}},
  \ and\ \bibinfo {author} {\bibfnamefont {B.}~\bibnamefont {Davidovitch}},\
  }\href@noop {} {\bibfield  {journal} {\bibinfo  {journal} {Journal of
  Elasticity}\ }\textbf {\bibinfo {volume} {119}},\ \bibinfo {pages} {137}
  (\bibinfo {year} {2014})}\BibitemShut {NoStop}%
\bibitem [{\citenamefont {Liang}\ and\ \citenamefont
  {Mahadevan}(2009)}]{liang2009shape}%
  \BibitemOpen
  \bibfield  {author} {\bibinfo {author} {\bibfnamefont {H.}~\bibnamefont
  {Liang}}\ and\ \bibinfo {author} {\bibfnamefont {L.}~\bibnamefont
  {Mahadevan}},\ }\href@noop {} {\bibfield  {journal} {\bibinfo  {journal}
  {Proceedings of the National Academy of Sciences}\ }\textbf {\bibinfo
  {volume} {106}},\ \bibinfo {pages} {22049} (\bibinfo {year}
  {2009})}\BibitemShut {NoStop}%
\bibitem [{\citenamefont {Audoly}\ and\ \citenamefont
  {Pomeau}(2010)}]{audoly2010elasticity}%
  \BibitemOpen
  \bibfield  {author} {\bibinfo {author} {\bibfnamefont {B.}~\bibnamefont
  {Audoly}}\ and\ \bibinfo {author} {\bibfnamefont {Y.}~\bibnamefont
  {Pomeau}},\ }\href@noop {} {\emph {\bibinfo {title} {Elasticity and geometry:
  from hair curls to the non-linear response of shells}}}\ (\bibinfo
  {publisher} {Oxford University Press Oxford},\ \bibinfo {year}
  {2010})\BibitemShut {NoStop}%
\bibitem [{\citenamefont {Baker~Jr}(1961)}]{baker1961application}%
  \BibitemOpen
  \bibfield  {author} {\bibinfo {author} {\bibfnamefont {G.~A.}\ \bibnamefont
  {Baker~Jr}},\ }\href@noop {} {\bibfield  {journal} {\bibinfo  {journal}
  {Physical Review}\ }\textbf {\bibinfo {volume} {124}},\ \bibinfo {pages}
  {768} (\bibinfo {year} {1961})}\BibitemShut {NoStop}%
\bibitem [{\citenamefont {Spivak}(1979{\natexlab{b}})}]{spivak}%
  \BibitemOpen
  \bibfield  {author} {\bibinfo {author} {\bibfnamefont {M.}~\bibnamefont
  {Spivak}},\ }\href@noop {} {\emph {\bibinfo {title} {A comprehensive
  introduction to differential geometry}}}\ (\bibinfo  {publisher} {Publish or
  Perish, Inc.},\ \bibinfo {address} {Berkeley},\ \bibinfo {year}
  {1979})\BibitemShut {NoStop}%
\bibitem [{\citenamefont {Lobkovsky}\ \emph {et~al.}(1995)\citenamefont
  {Lobkovsky}, \citenamefont {Gentges}, \citenamefont {Li}, \citenamefont
  {Morse},\ and\ \citenamefont {Witten}}]{science.paper}%
  \BibitemOpen
  \bibfield  {author} {\bibinfo {author} {\bibfnamefont {A.}~\bibnamefont
  {Lobkovsky}}, \bibinfo {author} {\bibfnamefont {S.}~\bibnamefont {Gentges}},
  \bibinfo {author} {\bibfnamefont {H.}~\bibnamefont {Li}}, \bibinfo {author}
  {\bibfnamefont {D.}~\bibnamefont {Morse}}, \ and\ \bibinfo {author}
  {\bibfnamefont {T.~A.}\ \bibnamefont {Witten}},\ }\href@noop {} {\bibfield
  {journal} {\bibinfo  {journal} {Science}\ }\textbf {\bibinfo {volume}
  {270}},\ \bibinfo {pages} {1482} (\bibinfo {year} {1995})}\BibitemShut
  {NoStop}%
\bibitem [{\citenamefont {Lobkovsky}(1996)}]{lobkovsky}%
  \BibitemOpen
  \bibfield  {author} {\bibinfo {author} {\bibfnamefont {A.~E.}\ \bibnamefont
  {Lobkovsky}},\ }\href@noop {} {\bibfield  {journal} {\bibinfo  {journal}
  {Phys. Rev. E.}\ }\textbf {\bibinfo {volume} {53}},\ \bibinfo {pages} {3750}
  (\bibinfo {year} {1996})}\BibitemShut {NoStop}%
\end{thebibliography}%

\end{document}